\documentclass[%
 reprint,
 amsmath,amssymb,
 aps,
]{revtex4-2}

\usepackage{graphicx} 
\usepackage{caption} 
\usepackage{times}
\usepackage{dcolumn}% Align table columns on decimal point
\usepackage{bm}% bold math
\usepackage{float} % Allows the forcing of figure location

\newcommand{\beginsupplement}{%
        \setcounter{table}{0}
        \renewcommand{\thetable}{S\arabic{table}} %
        \setcounter{figure}{0}
        \renewcommand{\thefigure}{S\arabic{figure}} %
        \setcounter{equation}{0}
        \renewcommand{\theequation}{S\arabic{equation}}
     }

\renewcommand{\epsilon}{\varepsilon}

\begin{document} 

\preprint{APS/123-QED}

\title{Optical assembly of nanostructures mediated by surface roughness} 

\author{Robert G. Felsted$^{1,2}$}
\author{Jaehun Chun$^{2,3\ast}$}
\author{Gregory K. Schenter$^2$}
\author{Alexander B. Bard$^1$}
\author{Xiaojing Xia$^4$}
\author{Peter J. Pauzauskie$^{2,5}$}
 \email{Corresponding Authors: peterpz@uw.edu, jaehun.chun@pnnl.gov}

\affiliation{$^1$Department of Chemistry, University of Washington, Seattle, Wa, 98195, USA}
\affiliation{$^2$Physical Sciences Division, Physical and Computational Sciences Directorate, Pacific Northwest National Laboratory, Richland, WA, 99352, USA}
\affiliation{$^3$Levich Institute and Department of Chemical Engineering, CUNY City College of New York, NY 10031, USA}
\affiliation{$^4$Department of Molecular Engineering and Science, University of Washington, Seattle, Wa, 98195, USA}
\affiliation{$^5$Materials Science and Engineering Department, University of Washington, Seattle, WA, 98195, USA}

\date{\today}

\begin{abstract}
Rigorous understanding of the self-assembly of colloidal nanocrystals is crucial to the development of tailored nanostructured materials. Despite extensive studies, a mechanistic understanding of self-assembly under non-equilibrium driven by an external field remains an ongoing challenge. We demonstrate self-assembly by optical tweezers imposing an external attractive field for cubic-phase sodium yttrium fluoride nanocrystals. We show that surface roughness of the nanocrystals is a decisive factor for contact leading to assembly between the nanocrystals, manifested by the roughness-dependent hydrodynamic resistivity. This provides direct evidence that dynamics are equally important to energetics in understanding self-assembly. These results have implications in a wide variety of different fields, such as in understanding the factors that mediate oriented attachment-based crystal growth or in interpreting the structure of binding sites on viruses.
\end{abstract}

%%%%%%%%%%%%%%%%% END OF PREAMBLE, START OF DOCUMENT %%%%%%%%%%%%%%%%

% Make the title.
\maketitle

\section*{Introduction}

Colloid-based materials have been of critical importance in various settings: fabrication and nanobiotechnology for sensors/actuators, energy storage, display, and printing\cite{Li2022,Zhang2010,Li2023,Hahm2022,Tan2018}, as well as crystal growth\cite{Mirabello2020,Liu2020}. Most of these areas involve careful and tailored control of self-assembly of nanoparticles, hence understanding the forces in self-assembly is integral to advances in such areas, as well as to fundamental colloidal science. Significant work has been done to understand assembly in colloidal systems, focusing on an energetic aspect of the interactions. DLVO theory\cite{Derjaguin1993,Verwey1947} has been commonly used to model the energetics of assembly and aggregation of colloids based on electrostatics and van der Waals interactions. However, in practice we cannot simply ignore the dynamic nature involved in self-assembly\cite{Lee2018,Nakouzi2018}; in fact, self-assembly results from a delicate balance between the energetics and dynamics of nanoparticles. This motivates the careful study of the hydrodynamics of colloids in order to reach a more complete understanding of the assembly process and the kinetics of particle assembly.

While the hydrodynamics of colloids can be studied in spontaneous self-assembly driven by Brownian motion, we can study the hydrodynamics more directly by inducing assembly with an external field. Many different types of fields can be used to achieve this effect, including electric, magnetic, temperature, and optical field gradients. Most of these can be easily applied to bulk solvent, but it can be challenging to apply local field gradients to individual colloidal particles without also influencing neighboring particles. This requires field effects to be focused sharply to extremely small volumes, usually on the scale of less than a few cubic micrometers for colloids. While this is challenging to do with most fields, optical forces can be focused to a small local effect through the use of optical tweezers\cite{Ashkin1970,ashkin1986}, which have an extremely wide set of applications\cite{Volpe2023, delic2020}. 

Here, we study assembly of cubic-phase sodium yttrium fluoride ($\alpha$-NAYF) crystals trapped in water with optical tweezers. Rare earth fluorides, such as $\alpha$-NAYF, are an interesting class of materials to study with optical tweezers due to their extremely wide range of applications in solar energy materials\cite{Shalav2005}, laser gain media\cite{Chou1986}, infrared upconversion phosphors\cite{Zhang2007,Fedotov2019,Kang2019}, vision therapy\cite{Ma2019}, gravity wave sensors\cite{Winstone2022}, and laser refrigeration\cite{ortiz2021,zhou2016,LuntzMartin2021}. We report the critical role of dynamics in assembly, revealed by optical tweezers driven assembly of $\alpha$-NAYF crystals, scaling analysis, and simulations of this process. In doing so, we reveal critical insights into the physical mechanisms that drive the assembly process of colloids.

\section*{Results and Discussion}

\begin{figure}
\includegraphics[width = 8 cm]{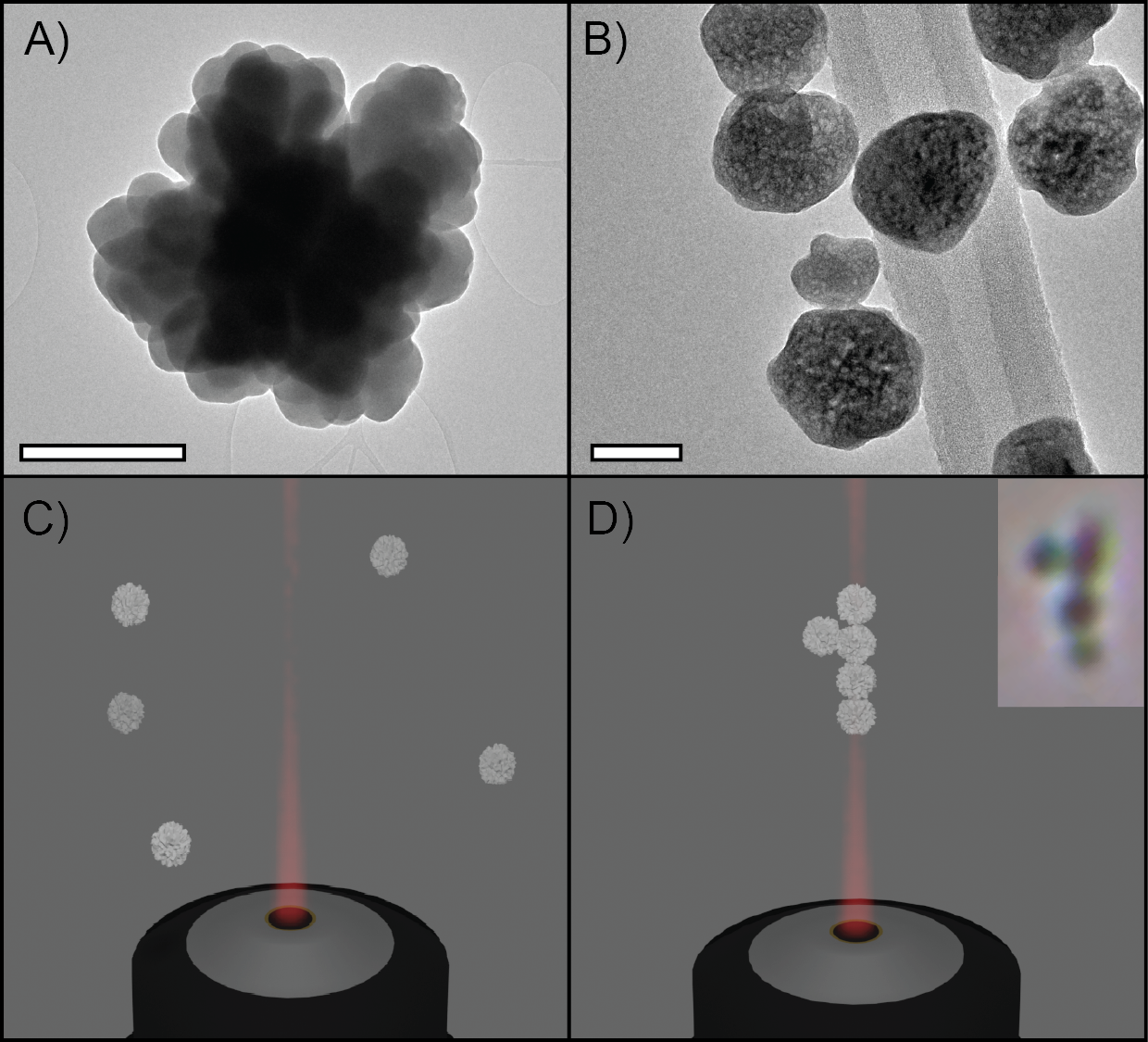}
\caption{(A) Transmission Electron Microscopy (TEM) image of the rough $\alpha$-NAYF used in the assembly experiments. Scale bar represents 400 nm. (B) TEM image of the smooth $\alpha$-NAYF grown for comparison in assembly experiments. Scale bar represents 50 nm. (C) Diagram of the assembly process. A 1020 nm laser formed an optical trap. Crystals near the focal plane are drawn inward due to the gradient force. (D) Diagram of an assembled chain of $\alpha$-NAYF crystals. An animation of the process can be seen in Supplementary Movie \ref{animation}. Inset: Micrograph image of the assembled chain as imaged through the trapping objective, taken from Supplementary Movie \ref{video}.}
\label{chain}
\end{figure}

We explore the optically driven attachment and assembly of $\alpha$-NAYF colloidal crystals into linear chains. A powder of $\alpha$-NAYF was synthesized hydrothermally using a modified synthesis from Felsted et al.\cite{Felsted2022} without organic ligands added, which resulted in a rough, quasi-spherical particle morphology (Fig. \ref{chain}A).  The rough $\alpha$-NAYF was suspended in water and investigated using optical tweezers, where they promptly assembled. In the presence of a strongly focused laser, the optical forces experienced by the $\alpha$-NAYF were enough to overcome repulsive forces. This brings the crystals close enough for the van der Waals force to become dominant, resulting in contact between crystals, shown in video in the Supplementary Movie \ref{video} and in print in Fig. \ref{chain}. Without the optical tweezers, we do not observe assembly in aqueous conditions. Smooth $\alpha$-NAYF crystals were also grown for comparison, seen in Fig. \ref{chain}B, but no assembly was ever observed with the smooth crystals, even with the use of optical tweezers.

Complex materials can be prepared using particle assembly, and this can be done using a variety of techniques. It is common to assemble particles using an active reaction or process, such as chemical reactions\cite{Hahm2022} or attachment to a substrate through some form of deposition\cite{Tan2018,Jin2018}. Recent work has also demonstrated wire assembly through a nanosoldering process\cite{crane2019}, which uses welding-like process to force assembly. In contrast with these chemical or thermal methods of forcing assembly, the observed assembly in our experiments occurs without any major chemical reaction or nanosoldering process. Instead, the fundamental properties of the $\alpha$-NAYF crystals are leveraged to direct assembly without inducing a chemical reaction or phase-change. The geometry of the superstructures we observe are consistent with those predicted in existing literature reports on theoretical calculations for optical forces on nonspherical nanostructures\cite{Borghese2008}, and this system shows a valuable physical realization of these principles while also suggesting further complexity within these systems to be characterized.

\subsection*{Forces for Modeling Energetics and Dynamics}

To understand the assembly process for crystals, we consider the Langevin equation with negligible particle inertia, called the generalized mobility relation. This describes a correlation between forces exerted on the particle and the resultant velocity: 
\begin{equation}
    \textbf{v} = \textbf{M}\cdot\Sigma \textbf{F}
\label{vel:force}
\end{equation}
where $\textbf{v}$ is the velocity of the particle, $\textbf{M}$ is the hydrodynamic mobility, and $\Sigma \textbf{F}$ is the sum of all forces acting on the particle. These forces include van der Waals, electrostatic, Brownian, and external forces. Considering these forces, Eq. \ref{vel:force} becomes:
\begin{equation}
    \textbf{v} = \textbf{M}\cdot(\textbf{F}_{vdw}+\textbf{F}_{elec}+\textbf{F}_B+\textbf{F}_{ext})
\label{velAllForces}
\end{equation}
where $\textbf{F}_{vdw}$ is the van der Waals force, $\textbf{F}_{elec}$ is the electrostatic force, $\textbf{F}_B$ is the Brownian force, and $\textbf{F}_{ext}$ is any external force imposed on the system. The attractive van der Waals force and the repulsive electrostatic force are conservative forces that can be derived from corresponding potential energies. The dynamic nature of the solvent results in non-conservative forces, including the random Brownian force and the dissipative hydrodynamic force (represented by the hydrodynamic mobility in Eq. \ref{velAllForces}) that are coupled via the fluctuation-dissipation theorem\cite{Russel1989}. The hydrodynamic force relevant to our system is purely repulsive,  hindering the assembly process; the hydrodynamic force becomes more repulsive as particles approach each other. This underscores the importance of accurately modeling the hydrodynamic forces when considering the forces at close separation distances that are decisive for the assembly process. All forces are dependent on material properties, such as dielectric permittivity and surface potential, as well as the system geometry (e.g. the particle size and separation between particles); more details on calculations of the forces can be found in supplementary materials\cite{Supple}.

The contribution from electrostatic forces is expected to be insignificant based on the small zeta potentials of the particles (approximately 30 mV, with no significant difference between the rough and smooth samples. See details in supplementary materials\cite{Supple}). This means the hydrodynamic force is expected to dominate the repulsive component of the forces between two $\alpha$-NAYF particles. This force arises from the relative motions between two particles, and it increases inversely with the separation distance between the maximum radii of particles ($h$)\cite{Russel1989}. This relationship only applies for separation distances much less than the radius of particle ($a$), also known as the lubrication regime (at $h \ll a$, schematic shown in Fig. \ref{Spheres}). However, this model assumes a perfectly spherical particle, which is not consistent with the morphology of many real-world particles, notably including the $\alpha$-NAYF particles we used. Outside the lubrication regime, hydrodynamic interactions between two particles become less significant, and the overall force approaches the Stokes’ drag force (6$\pi \mu av$)\cite{Jeffrey1984}. Here, $\mu$ denotes the viscosity of the surrounding fluid and $v$ is the magnitude of the particle velocity. Previous studies\cite{Jeffrey1984,Kim1991} clearly demonstrated that the hydrodynamic force along the line of centers of the particles is the most significant (scaled as $O(1/\epsilon)$), in contrast to the perpendicular direction, which scales as $O(\rm{log}(1/\epsilon)))$, where $\epsilon$ is a normalized separation distance between spheres by $a$. For this reason, our hydrodynamic modeling focused on the effect along the center line of the particles.

Using the Oseen tensor and a scheme to generate surface meshes, we used the BEST\cite{Aragon2004,Aragon2006} software to calculate hydrodynamic forces as a function of separation distance for rough particles, as detailed in the supplementary materials\cite{Supple}. Fig. \ref{forces} shows a normalized hydrodynamic resistivity by 6$\pi \mu a$, which is directly proportional to the hydrodynamic force between two particles for smooth and rough sphere cases. For rough spheres, a finite hydrodynamic repulsive force (about 2 times the Stokes’ drag force) is manifested at $\epsilon \ll 1$ due to the surface roughness that reasonably represents the synthesized rough $\alpha$-NAYF. This is consistent with the lubrication analysis by Jenkins \& Koenders\cite{Jenkins2005} that suggested the existence of $O(1)$ hydrodynamic resistivity between two rough spherical particles along the line of centers at $\epsilon \ll 1$. More importantly, the finite hydrodynamic resistivity is over an order of magnitude smaller than both numerical solutions and the asymptotic formula at the lubrication limit for a smooth sphere by Jeffrey \& Onishi\cite{Jeffrey1984}, represented by $O(1/\epsilon)$. The extremely weak separation dependence and lack of appreciable hydrodynamic interactions provide a reasonable rationale to ignore hydrodynamic interactions between rough spheres and instead use the Stokes' drag force. This implies that the hydrodynamic interactions perpendicular to the line of centers between the particles can also be ignored, while one needs to employ appreciable hydrodynamic interactions for smooth spherical particles in both directions, shown by Jeffrey \& Onishi\cite{Jeffrey1984}. Therefore, our calculations imply that sufficient surface roughness makes particles mobile at close separation distances, allowing the optical tweezers to be more effective at driving rough particles together by minimizing the effect of repulsive hydrodynamic interactions on particle assembly. This emphasizes the critical role of hydrodynamics in understanding particle assembly and how mitigating hydrodynamic resistivity can increase the probability and rate of particle assembly.

\begin{figure}
\includegraphics[width = 8cm]{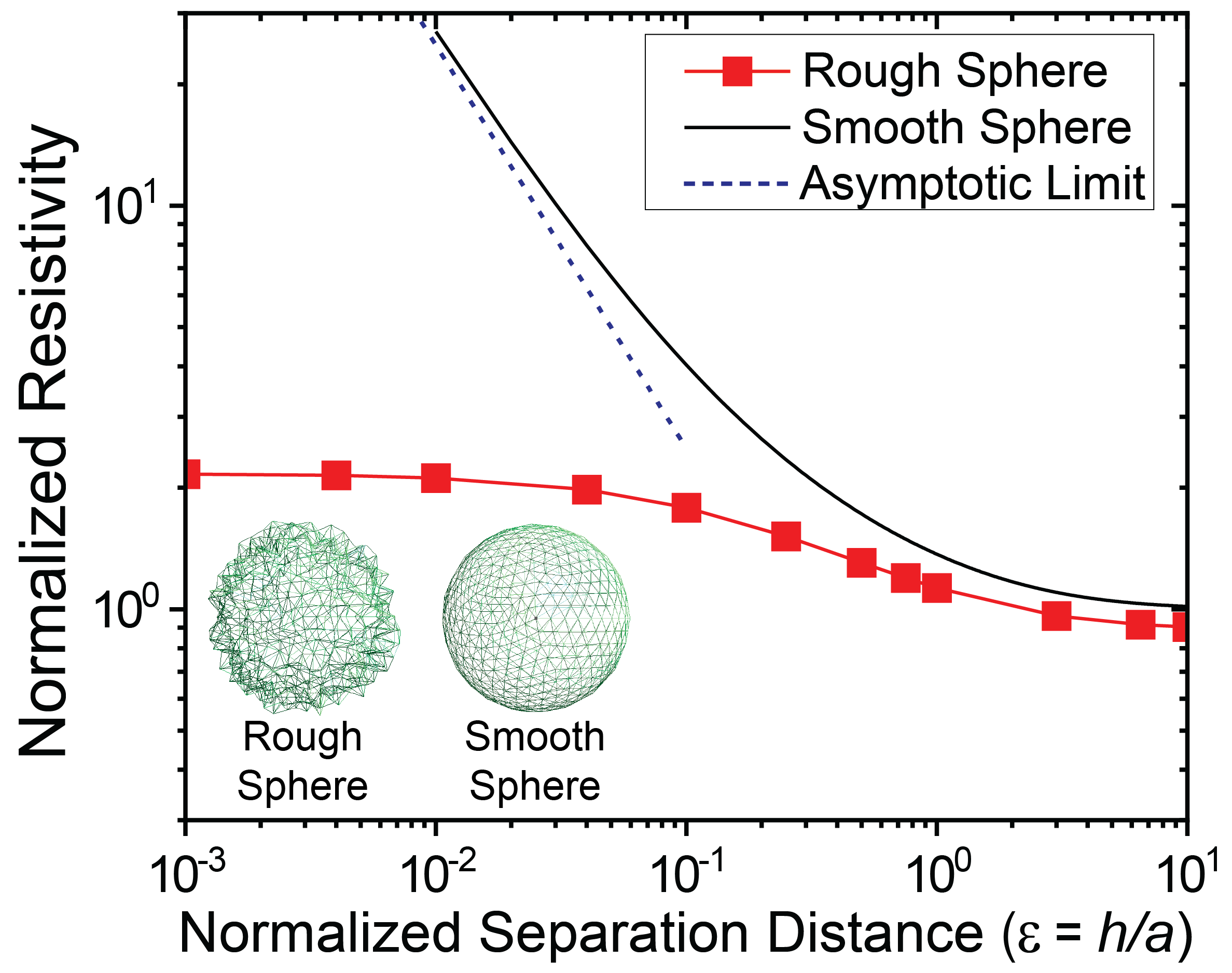}
\caption{Comparison of hydrodynamic resistivity (normalized by $6\pi \mu a$) between two spheres in the cases of smooth spheres versus rough spheres. The asymptotic limit for two smooth spheres comes from the analytical solution presented in Jeffrey \& Onishi\cite{Jeffrey1984}. The hydrodynamic resistivity for the rough sphere also approaches an asymptotic value. Details on the calculations can be found in the supplementary materials\cite{Supple}. Inset: Models of the rough spheres versus smooth spheres used for the hydrodynamic resistivity simulation.}
\label{forces}
\end{figure}

The last major force component to consider is the external force caused by the optical tweezers. This provides a net attractive force which is critical to the observed assembly, and details about the calculation of the optical force can be found in the supplementary materials\cite{Supple}. This force acts in opposition to the hydrodynamic resistivity and electrostatic repulsion, allowing the particles to be close enough for van der Waals interactions to occur.

\subsection*{Langevin Dynamics Simulation for Self-Assembly}

\begin{figure*}
\includegraphics[width = 12cm]{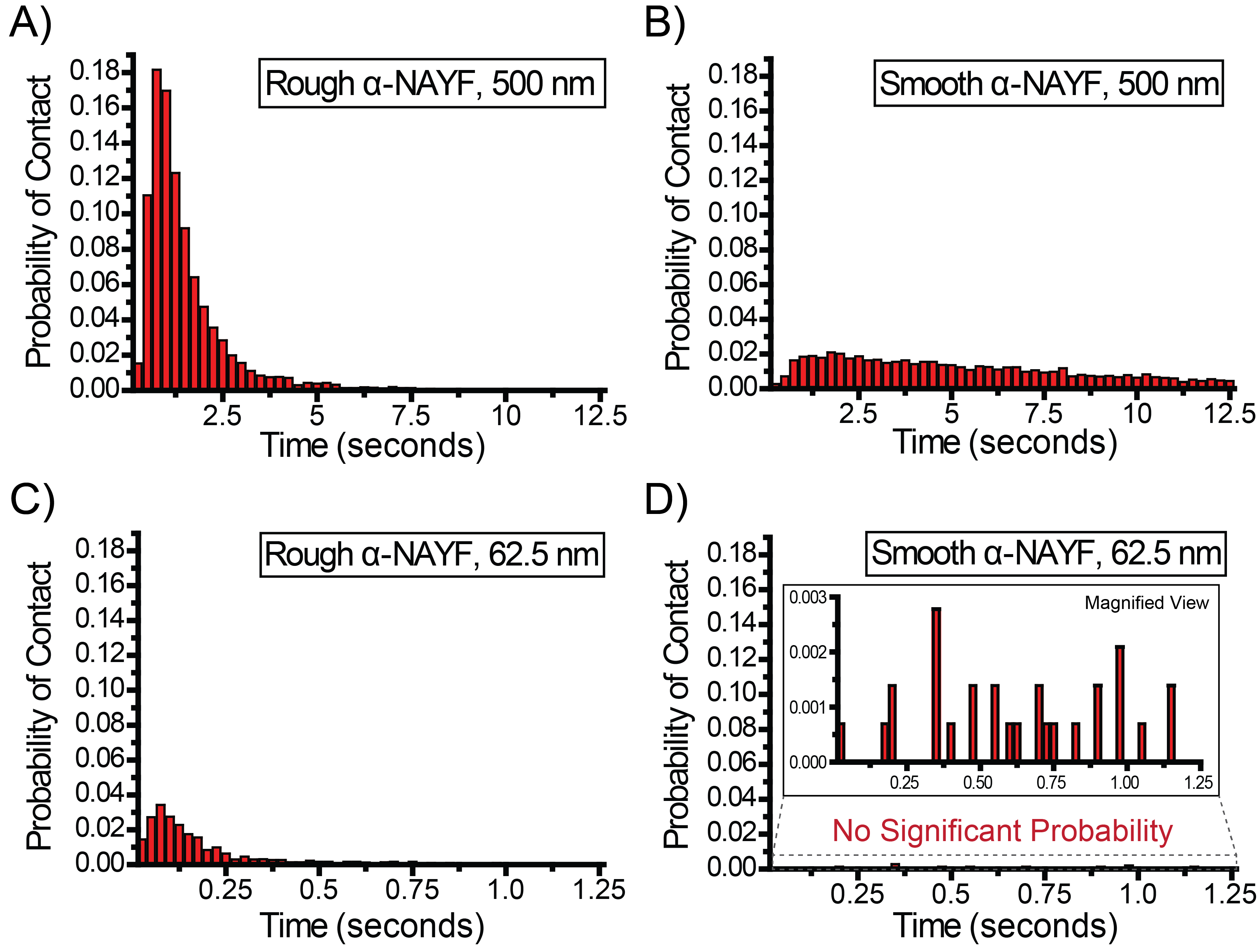}
\caption{Calculated probability density functions for the contact of two $\alpha$-NAYF crystals inside an optical trap. All graphs share the same vertical axis scaling to allow for easy comparison. Separate cases are shown for (A) Rough $\alpha$-NAYF with a radius of 500 nm, (B) Smooth $\alpha$-NAYF with a radius of 500 nm, (C) Rough $\alpha$-NAYF with a radius of 62.5 nm, and (D) Smooth $\alpha$-NAYF with a radius of 62.5 nm. (A) and (D) represent the measured materials in the experiments reported here while (B) and (C) are calculated to compare the effect of size on the probability of contact between two crystals. Time scales were chosen to allow for convergence, and enough time steps were binned to guarantee that the elapsed time for the binned events well exceeded the viscous relaxation time, making sure no contact events were missed. No significant probability of contact was found in the case of small, smooth $\alpha$-NAYF as shown in (D).}
\label{pdfs}
\end{figure*}

Using the forces described above and detailed parameters shown in the supplementary materials\cite{Supple}, we performed Langevin dynamic simulations to examine if two $\alpha$-NAYF crystals in a set of optical tweezers would reliably come into contact with each other, leading to the assembly. From the simulations, collision times are binned and averaged over an ensemble of 8000 instances, represented as a probability density function for the collision time, shown in Fig. \ref{pdfs}. The probability density functions clearly show that a collision is highly probable for rough crystals (Fig. \ref{pdfs}A and \ref{pdfs}C) with all simulated crystals colliding within a few seconds, but the probability of contact for smooth crystals is severely reduced and virtually unseen for the sizes of smooth $\alpha$-NAYF grown experimentally (Fig. \ref{pdfs}B and \ref{pdfs}D). The results match the experimentally-observed events in our optical tweezers, suggesting that the roughness of particles directly affects the probability of physical contact between two particles. This agrees with the order of magnitude difference in the hydrodynamic resistivity at close proximities (Fig. \ref{forces}). The probability also increases as the size of the crystal increases. Note that, due to synthesis limitations, we are unable to grow smooth and rough-NAYF at identical sizes. Because of this limitation, we ran simulations for smooth and rough-NAYF at sizes they cannot normally be grown to control for size effects in collision probability, while keeping other parameters constant in the simulation. 

Further physical insights on assembly can be obtained by considering a coordinated interplay between the two major attractive forces in this system: the optical gradient force ($\textbf{F}_{grad}$) due to optical tweezers and the van der Waals force ($\textbf{F}_{vdw}$). The interplay represents the gradient force bringing particles closer together until the van der Waals force becomes dominant. Based on the dipole approximation and assuming a low separation distance ($h/\omega_0 \ll 1$) where $\omega_0$ is the beam waist, the magnitude of $\textbf{F}_{grad}$ in radial direction can be approximated by
\begin{equation}
    F_{grad} = -\frac{2\alpha'_pI_0h}{cn_m\omega_0^2}
    \label{fgrad_eq}
\end{equation}
where $\alpha'_p$ is a real part of the polarizability of the particle, $I_0$ is the maximum laser intensity, $c$ is the speed of light, and $n_m$ is the refractive index of the surrounding medium\cite{Pesce2020}. On the other hand, $\textit{F}_{vdw}$ (=$|\textbf{F}_{vdw}|$) in the radial direction for identical spherical particles at the similar separation distances ($h/a \ll 1$) is described by 
\begin{equation}
    F_{vdw} = -\frac{A_Ha}{h^2}
    \label{vdw_equation}
\end{equation}
where $A_H$ is the Hamaker constant that can be calculated from the frequency-dependence of the dielectric functions of each constituent of the system, representing the strength of van der Waals interaction between the particles with the surrounding medium\cite{Russel1989,Israel6,Israel13}. 

To investigate the interplay of $F_{vdw}$ with $F_{grad}$, we will assume that, if there is a particle at the center of the optical trap, Eq. \ref{fgrad_eq} will still apply to a nearby particle that is not yet in the trap. Future work will address more complex optical forces, such as optical binding\cite{Burns1989,Dholakia2010,Donato2019,Forbes2019}, which are challenging to calculate for this system due to the nontrivial morphology of the rough particles. Assuming that a particle is already at the tweezers' center, we can define a cross-over separation distance, $\hat{h}$, from a competition between the magnitude of the two attractive forces $\textit{F}_{grad}$ and $\textit{F}_{vdw}$. This competition leads to a cross-over separation distance of
\begin{equation}
    \hat{h} = \left(\frac{A_Ha}{\kappa_r}\right)^{1/3}
\end{equation}
where $\kappa_r \equiv 2\alpha'_pI_0/cn_m\omega_0^2$. This cross-over distance is the separation distance below which the van der Waals contribution to the attractive forces becomes the dominant force. Because the van der Waals force becomes noticeable at $\hat{h} \leq a$ in general, $\hat{h} \leq a$ would be a necessary condition for the optical field driven attachment to take place. Therefore, the maximum distance we would expect van der Waals force to overtake the optical gradient force and begin driving assembly would be at the limit of $\hat{h} = a$. This leads to a minimum electric field intensity enabling the attachment,
\begin{equation}
    (I_0)_{min} \approx \frac{A_Hcn_m\omega_0^2}{2\alpha'_pa^2}
\end{equation}
for a given system. The simple scaling indicates that one needs to tailor both $\omega_0$ and $a$ to drive the attachment for a given system. 

\subsection*{Implication of Hydrodynamic Mobility on Assembly Kinetics}

While the above physical rationale provides the feasibility of the coordinated interplay between two attractive forces leading to contact, it is insufficient to understand kinetics of assembly because it does not directly correspond to an attachment frequency. This can be understood by considering a generalized relation between velocity and force, as seen in Eq. \ref{vel:force}. This can be generally conceived as a scalar-based generalized mobility relation $v = MF$ where $M$ and $F$ denote the magnitudes of hydrodynamic mobility and force between particles, respectively. The relation indicates that the relative motion between the particles (involving a time scale for the attachment) is heavily influenced by the hydrodynamic mobility. Considering particle contact, the most appropriate attractive force for predicting the kinetics would be $F_{vdw}$ due its dominance at short separation distances. Per Eq. \ref{vdw_equation}, $F_{vdw}$ scales by $O(\epsilon^{-2})$. The mobility scales as $M_{rough} \sim O(1)$ for the rough particle and $M_{smooth} \sim O(\epsilon)$ for the smooth particle in the lubrication regime, $\epsilon \ll O(1)$ (Fig. \ref{forces}). This leads to different scaling in the relative particle velocity for rough and smooth cases; $v_{rough} \sim O(\epsilon^{-2})$ and $v_{smooth} \sim O(\varepsilon^{-1})$. This implies that a time scale for approaching motion for the rough particle would be an order of magnitude faster than that for the smooth particle, leading to the noticeable difference in probability of attachment events (Fig. \ref{pdfs}).

\section*{Conclusions}

We have observed the optically-driven assembly of $\alpha$-NAYF single crystals into a linear chains that persist for an indeterminately long period of time. This assembly is counter-intuitive, yet can be reliably induced through the use of optical tweezers for rough $\alpha$-NAYF crystals. Simulations show that the assembly is enabled by the surface roughness of the $\alpha$-NAYF crystals, which directly reduces the hydrodynamic repulsive force experienced when two colloidal crystals come into close proximity. The presence of surface roughness enhances the probability of particles coming into physical contact, enabling assembly.

The coordinated interplay between $F_{grad}$ and $F_{vdw}$ brings a unique system to reveal the importance of hydrodynamics on particle assembly, which is made possible by the use of optical tweezers. An analogous case has been observed in the coagulation of aerosols under turbulence where the coordinated interplay between $F_{shear}$ (turbulent shear force) and $F_{vdw}$ takes place. In this case, the non-continuum nature of gas causes a significant variation of the hydrodynamic forces\cite{Chun2005}; the variation in the hydrodynamic force results in a drastic difference for the coagulation rate of aerosols.

The roughness of the surface causes the hydrodynamic forces to be significantly reduced and effectively invariant over all separation distances, increasing the probability of direct contact between two nanoparticles in solution. This has major impacts on any system that involves the contact of two bodies in an aqueous system, and could potentially have explanatory power in systems such as the dynamics underlying oriented attachment crystal growth or the morphologies of proteins on the surfaces of viruses and microorganisms. When used in conjunction with 3D tracking\cite{Johnson2022}, the exact dynamics of complex systems could be mapped out in real time. The attachment method itself has direct applications in additive manufacturing at the micro and nano size scales and in the production of, for example, complex anti-counterfeiting structures. In addition to these potential applications, the use of optical tweezers allows experiments to isolate individual forces acting on colloidal particles. This allows us to probe each force directly and allows for the precise analysis of the hydrodynamic force directly, demonstrating its unique nature and its critical influence over colloidal aggregation and assembly.

\subsection*{Acknowledgments}

R.G.F. and P.J.P. gratefully acknowledge financial support from the ONR-BRC Grant N00014-18-1-2370, which supported the synthesis and design of materials. X.X. and P.J.P. acknowledge support from the MURI:MARBLe project under the auspices of the Air Force Office of Scientific Research (Award No. FA9550-16-1-0362), which supported optical trapping efforts. Transmission electron microscopy of materials was supported by the U.S. Department of Energy, Office of Science (DOE) Basic Energy Sciences (BES) Synthesis and Processing Science. The development of the Langevin dynamic and hydrodynamic simulations was partially supported by DOE/BES Chemical Sciences, Geosciences, and Biosciences Division, Chemical Physics and Interfacial Sciences Program, FWP 16249. Theories for all the simulations and analyses were supported by DOE/BES, Synthesis and Processing Science Program, FWP 67554. Sample characterization was conducted at the University of Washington Molecular Analysis Facility, which is supported in part by the National Science Foundation (Grant No. ECC-1542101), the University of Washington, the Molecular Engineering \& Sciences Institute, the Clean Energy Institute, and the National Institutes of Health.

%R.G.F. and P.J.P. conceived of and designed the experiment. R.G.F. and X.X. collected video footage of the assembly process. R.G.F. and A.B.B. performed electron microscopy and imaging of the materials. R.G.F. performed data analysis. J.C. developed the theories, simulations, and analyses for the colloidal forces. G.K.S. developed the simulation scheme and subsequent analysis for particle collision. R.G.F. and J.C. wrote the manuscript, and R.G.F., P.J.P., J.C., and G.K.S. edited the manuscript.

\bibliography{Biblio}

%%% NOTE TO SELF: "C. Johnson in Nature Methods 2022." is currently the last reference in the main text.

%\section*{List of Supplementary Materials:}
%Materials and Methods \\
%Supplementary Text \\
%Equations S1 to S6 \\
%Figures S1 to S8 \\
%Supplementary Movies S1 to S2 \\
%References(48-62)

\clearpage

\setcounter{page}{1}

% ADD SI TEXT TO DOCUMENT
\beginsupplement
\begin{widetext}
\begin{center}

{\LARGE Supplementary Materials for}\\

\hfill

{\Large \textbf{Optical assembly of nanostructures via surface
roughness}}\\

\hfill

{Robert G. Felsted,$^{1,2}$ Jaehun Chun,$^{2,3\ast}$ Gregory K. Schenter,$^{2}$ Alexander B. Bard,$^{1}$ \\Xiaojing Xia,$^{4}$ Peter J. Pauzauskie$^{2,5\ast}$\\

\hfill

\normalsize{$^{1}$Department of Chemistry, University of Washington, Seattle, WA, 98195, USA}\\
\normalsize{$^{2}$Physical Sciences Division, Physical and Computational Sciences Directorate,}\\
\normalsize{Pacific Northwest National Laboratory, Richland, WA, 99352, USA}\\
\normalsize{$^{3}$Levich Institute and Department of Chemical Engineering, CUNY City College of New York,}\\
\normalsize{NY 10031, USA}\\
\normalsize{$^{4}$Depart of Molecular Engineering and Science, University of Washington, Seattle, Wa, 98125, USA}\\
\normalsize{$^{5}$Materials Science and Engineering Department, University of Washington,}\\
\normalsize{Seattle, WA, 98195, USA}\\

\hfill
 
{\small $^\ast$To whom correspondence should be addressed; E-mail: peterpz@uw.edu, jaehun.chun@pnnl.gov}
}

\end{center}

\noindent \textbf{This PDF file includes:} \\
Materials and Methods \\
Supplementary Text \\
Equations S1 to S6 \\
Figures S1 to S8 \\
Supplementary Movies S1 to S2 \\
References(46-52)

\clearpage
\end{widetext}
\section*{Materials and Methods}
The $\alpha$-NAYF was synthesized under hydrothermal conditions. 7 mL of a 0.2 M YCl$_3$:10\% YbCl$_3$ (99.9\% pure YCl$_3$, 99.999\% pure YbCl$_3$, trace metal bases for both) aqueous solution was thoroughly mixed with 7 mL of a 0.8 M NaF (99.5\% pure) aqueous solution. This mixture was transferred to an autoclave liner. The autoclave (Parr Instrument Company, 4747) was then assembled and heated at 100°C for 24 hours. The resulting product was centrifuged and then washed twice with water and twice with ethanol. The $\alpha$-NAYF product was dried at 70°C overnight to remove any residual ethanol, and chemical composition was confirmed through XRD analysis. The reaction is independent of reaction volume and has been successfully scaled up to a 1 L reaction volume with no adverse effects.

\begin{figure*}
\includegraphics[width = 16cm]{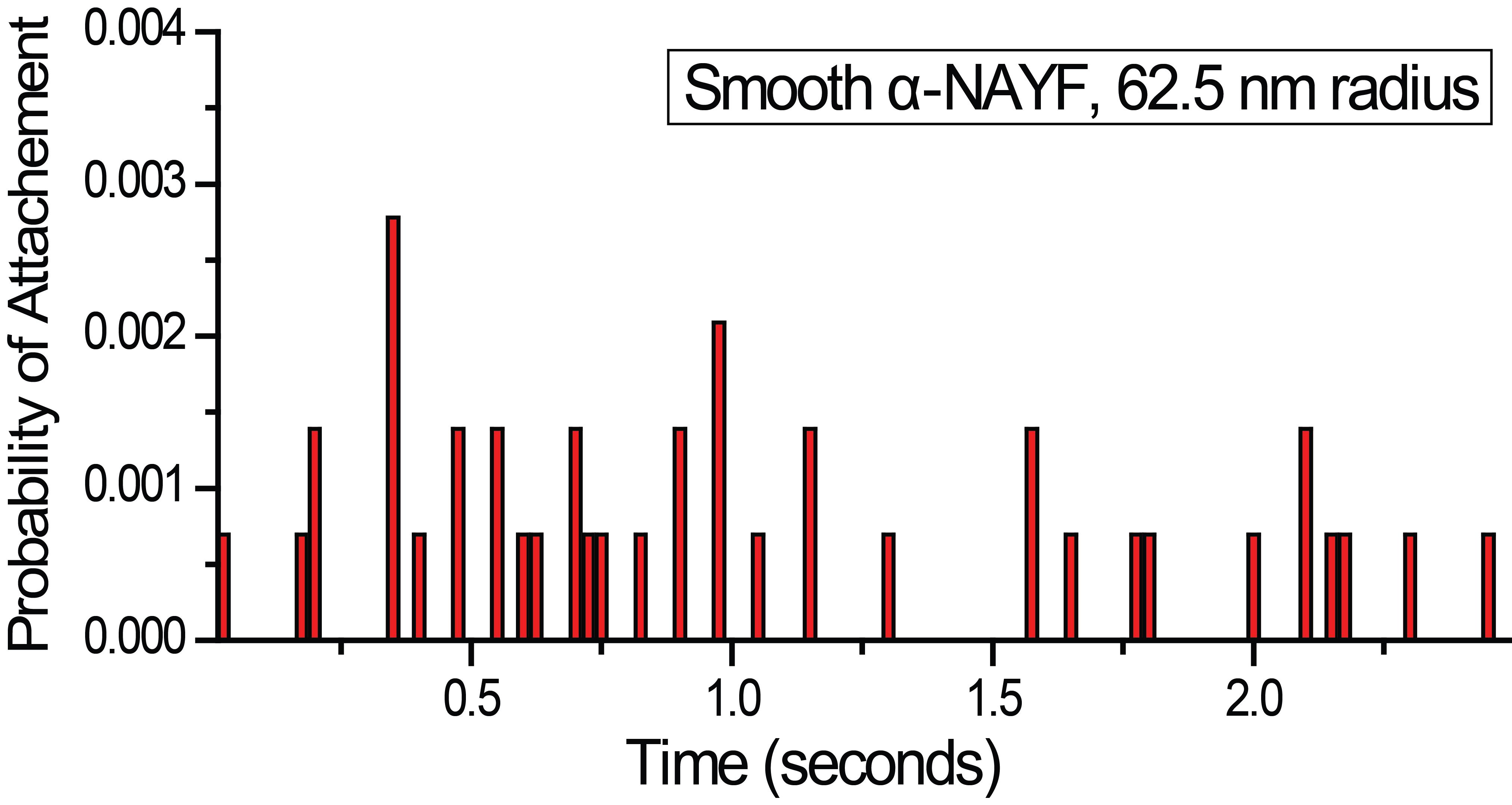}
\caption{A re-scaled plot from section (D) in figure \ref{pdfs}. The time axis has been extended, and the probability axis has been re-scaled to be more visible. The probability does not form to a curve at timescales relevant to these experiments. Of note, the probabilities shown are all integer multiples of the lowest (and most common) probability plotted, suggesting that most consist of one or two runs where contact randomly occurred. This can be considered insignificant and is not relevant to forming a understanding of the probability density and kinetics of assembly, as noted in figure \ref{pdfs}.} 
\label{Small}
\end{figure*}

To form the smooth $\alpha$-NAYF, the synthesis was identical to the process above, except that the 0.2 M YCl$_3$:10\% YbCl$_3$ solution was not dissolved into pure water, but rather was prepared using a solution of 0.2 M ethylenediaminetetraacetic acid (EDTA) that had been neutralized to a pH of 7.0. This mixture allowed the YCl$_3$ and YbCl$_3$ to be chelated to the EDTA, and the EDTA acted as a ligand in the final product. The smooth $\alpha$-NAYF morphology can be seen in Fig. \ref{Small}.

\begin{figure*}
\includegraphics[width = 16cm]{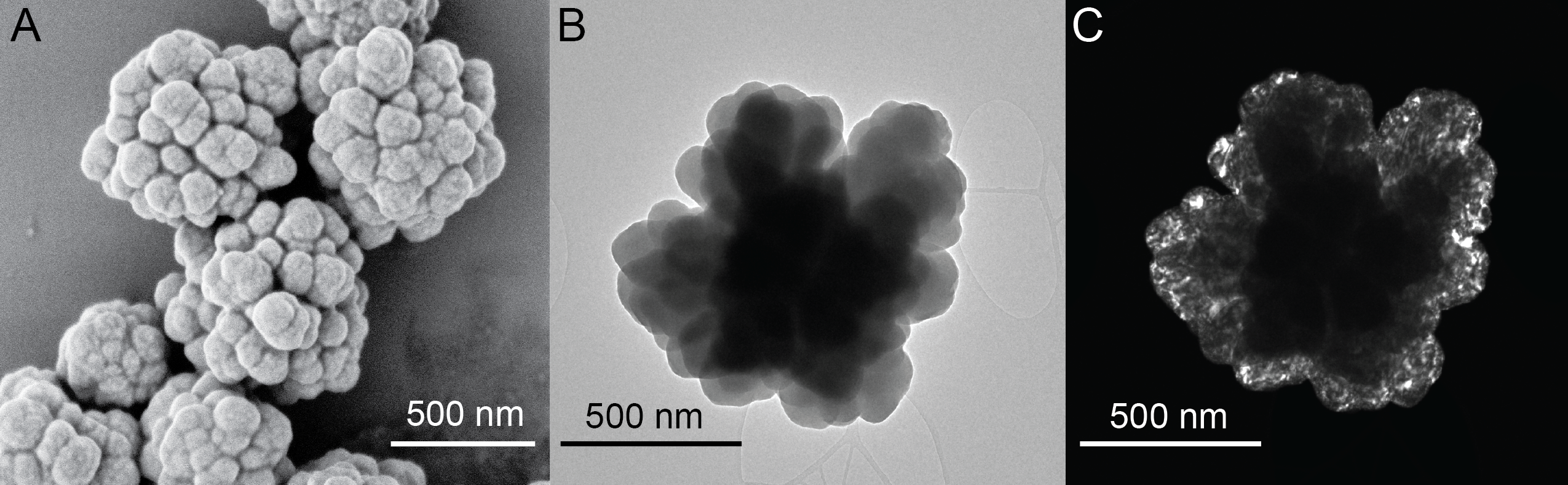}
\caption{(A) SEM image of $\alpha$-NAYF used for assembly experiments. The rough surface of the crystals is unique to the hydrothermal synthesis run without added ligands, such as oleic acid, ethylenediaminetetraacetic acid, or citric acid. (B) TEM bright field micrograph of a single $\alpha$-NAYF crystal. (C) Dark field micrograph from electron scattering from a single crystallographic point in a select area electron diffraction scan, indicating the portions of the image that diffract to the same point and are therefore part of the same single crystal. Reduced scattering is seen from the center of the crystal due to increased thickness of the scattering material.}
\label{NAYF}
\end{figure*}

\begin{figure*}
\includegraphics[width = 12cm]{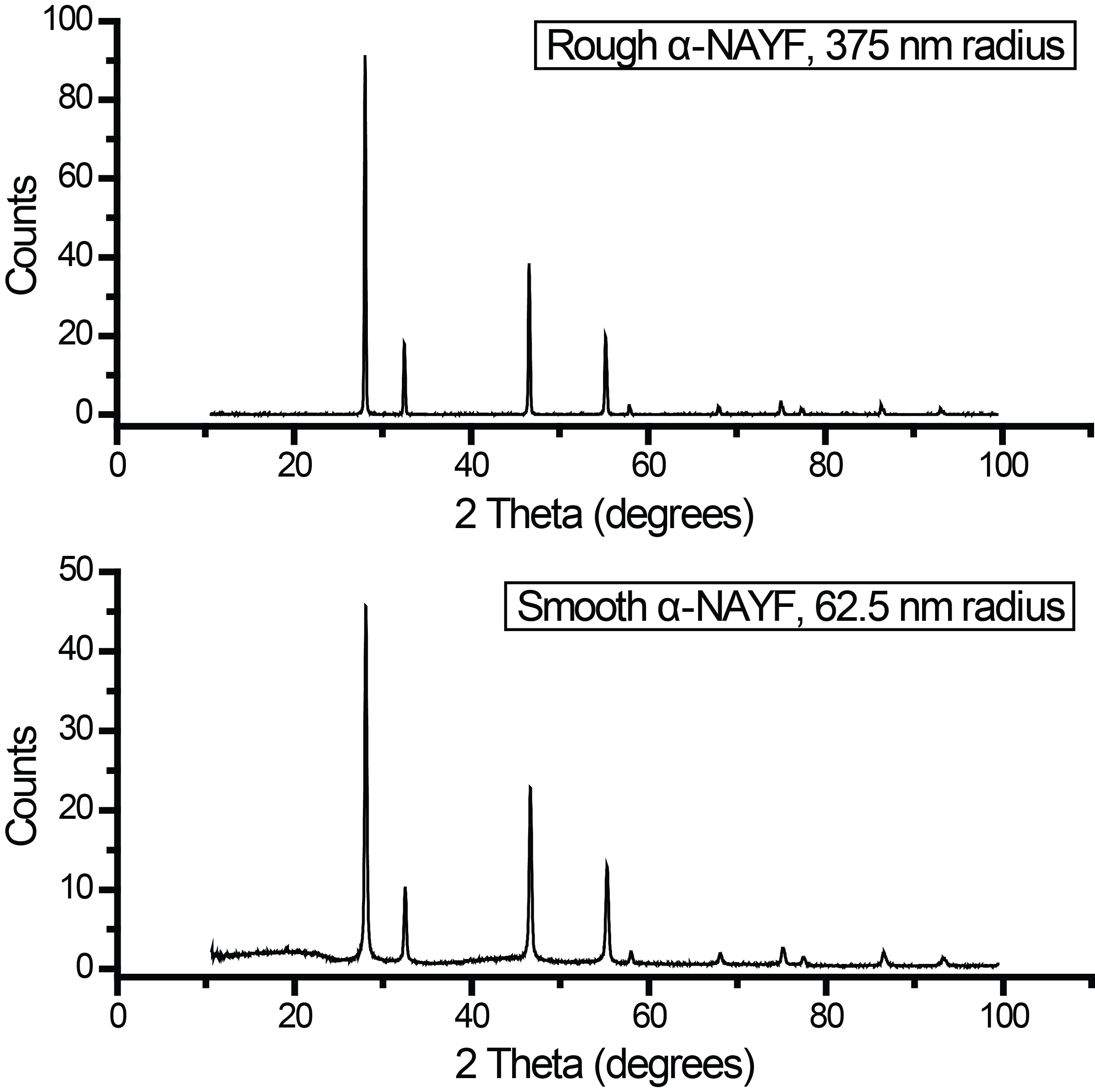}
\centering
\caption{X-ray diffractrometry spectra showing identical composition for both the rough and smooth $\alpha$-NAYF crystals. Some broadness can be seen in the peaks of the smooth $\alpha$-NAYF due to Scherrer broadening at crystal sizes seen in the smooth case. Other than this broadening, the peaks are functionally identical, showing that the materials are the same crystal phase.}
\label{XRD}
\end{figure*}

The result of the synthesis was an unusual morphology of $\alpha$-NAYF that was very rough, seen in figure \ref{NAYF}A. To determine if the rough crystals were single crystalline, we used TEM analysis and electron diffraction. Dark field TEM data in figure \ref{NAYF}B and C show that the rough $\alpha$-NAYF particles are single crystals despite their rough morphology. Both the rough and smooth $\alpha$-NAYF crystals were confirmed to be the same crystalline phase using XRD shown in figure \ref{XRD}. 

The $\alpha$-NAYF crystals were studied in a home-built single-beam optical tweezers instrument using a 1020 nm laser for the trapping beam (QPhotonics QFBGLD-1020-400) and a 100x magnification objective with a numerical aperture of 1.25. Images and video of the assembly process were collected using a ThorLabs scientific camera. Colloids were prepared by suspending milligram scale quantities (depending on exact desired concentration) of $\alpha$-NAYF powder in 1 mL of pure water. This mixture was sonicated for 15 minutes and mixed using a vortex mixer to ensure the dry crystals mixed well with the water to form the colloid.

When a single $\alpha$-NAYF crystal was trapped, no change in the crystal was observed, and the crystal remained stable in the trap until the trap was released. However, when multiple rough $\alpha$-NAYF crystals were introduced into the trap, the crystals unexpectedly assembled into a linear chain aligned with the trapping beam, shown in video form in Supplementary Movie \ref{video} and in print in figure \ref{chain}. The linear structure persisted after the trapping beam was removed and no separation of assembled crystals was observed. Spontaneous assembly without the trapping beam was never observed, though occasional dimers of two attached $\alpha$-NAYF crystals could be found immediately after suspension, which may only be the result of aggregation or twinning during the synthesis process. Chain lengths of up to 10 $\alpha$-NAYF crystals (approximately 8 $\mu$m long) have been made, with the chains eventually destabilizing as they increase in length. One potential limit for the chain length may be from the focal length of the laser used in the optical tweezers. After the chain exceeded 10 crystals, the far end of the chain rested well outside the focal plane of the optical trap and the linear structure began to be unstable in the trap as most of the mass was outside the region where the optical force was the strongest. If the crystals were anchored to a surface or if higher optical power was used to form a sufficiently more stable trap for long chains, it may be possible to make longer chains and structures.

The optical forces were simulated using the Optical Tweezers Toolbox, which is capable of modeling the forces on nanoparticles in an optical trap\cite{Nieminen2007}. The smooth $\alpha$-NAYF crystals were simulated using a sphere, while the rough $\alpha$-NAYF crystals were simulated using a 3D model of a rough sphere generated through a method described below. These different shapes were simulated to extract out the optical forces on them at every position in space. These forces were then exported and used for simulating particle collision for two crystals entering an optical trap.

\section*{Supplementary Text}

\subsection*{Descriptions of Brownian, Optical, Electrostatic, and van der Waals Forces}

For colloidal particles in the presence of optical tweezers, many forces simultaneously influence the velocity (thus position) of each particle.

\subsubsection*{Brownian Forces}

Random Brownian forces ($\textbf{F}_B$) on the particle at a given time $t$ can be obtained from $\langle\textbf{F}_B(t)\textbf{F}_B(0)\rangle =12\pi\mu akT\textbf{K}\delta(t)$, along with $\langle \textbf{F}_B\rangle=0$, where $k$ is the Boltzmann constant, $T$ is the system temperature, $\mu$ is the viscosity of surrounding fluid, $a$ is the particle radius, and $\delta$ is the delta function\cite{Russel1989}. $\textbf{K}$ is a normalized hydrodynamic resistivity tensor by $6\pi\mu a$, relating a hydrodynamic force on a particle due to its velocity (or the inverse of a corresponding normalized hydrodynamic mobility tensor $\textbf{M}$ by $1/6\pi\mu a$). It is important to note that $\textbf{K}$ and $\textbf{M}$ are dependent on a separation between two particles, $h$. Following the work by Jeffrey and Onishi\cite{Jeffrey1984}, $\textbf{K}(h)=A(h) \textbf{ee} + B(h) (\textbf{I}-\textbf{ee})$ where $A$ and $B$ are resistivity functions along the line of centers of particles and perpendicular to the line of centers, respectively. Here, $\textbf{I}$ and $\textbf{e}$ denote the identity tensor and unit vector along the line of centers. Therefore, an x-component of the Brownian force, $F_{B,x}$, can be obtained from $\sqrt{(12\pi\mu akT(Ae_xe_x+B(1-e_xe_x))/\Delta t)}\cdot N$ where $N$ is a Gaussian random number with a mean of zero and unity standard deviation, $\Delta t$ is a relevant time step, and $e_x$ is a x-component of $\textbf{e}$. Similarly, y- and z- components of the Brownian force on the particle can be calculated.

\subsubsection*{Optical Forces}

The presence of optical tweezers introduces three forms of optical force on particles in the tweezers: the gradient force, the scattering force (also known as radiation pressure), and the spin-curl force\cite{Pesce2020}. The spin-curl force is only present with non-homogeneous polarization, which does not apply to our experimental conditions. The scattering force is always present with optical tweezers, but the direction of the scattering is in the direction of the beam propagation of the trapping laser. As such, it does not apply a significant relative attractive or repulsive force between particles when considering approach at close separation distances and especially if considering approach from a radial direction. The main force of interest for the experiments shown here is the gradient force, which draws particles into the focus of the trapping laser. For simplicity of modeling the gradient force, we apply the dipole approximation which treats the trapped particles as dipoles with homogeneous internal fields. If this approximation is used, then the magnitude of time-averaged gradient force can be expressed as:
\begin{equation}
    |\textbf{F}_{grad}(\textbf{r})| = \frac{1}{4}\alpha_p'\nabla |\textbf{E}_i(\textbf{r})|^2
\end{equation}
where $r$ is the radial distance from the trap, $\alpha_p'$ is the real part of the polarizability of the particle in the trap, and $\textbf{E}_i$ is the electric field incident on the particle from the laser. This can be simplified further using the relation between the electric field and the intensity of the electric field for a Gaussian beam, which yields
\begin{equation}
    |\textbf{F}_{grad}(\textbf{r})| = \frac{1}{2}\frac{\alpha_p'}{cn_m}\nabla I_i(\textbf{r})
\end{equation}
where $c$ is the speed of light, $n_m$ is the refractive index of the surrounding medium, and $I_i$ is the intensity of the electric field incident on the particle. This equation for $\textbf{F}_{grad}$ is used in our calculations to determine the strength of the attractive external force ($\textbf{F}_{ext}$) on the particles.

These forces were simulated using the Optical Tweezers Toolbox, which is capable of modeling the forces on nanoparticles in an optical trap\cite{Nieminen2007}. The smooth $\alpha$-NAYF crystals were simulated using a sphere, while the rough $\alpha$-NAYF crystals were simulated using the 3D model of a rough sphere shown in Fig. \ref{forces}. These different shapes were simulated to extract out the optical forces on them at every position in space. These forces were then exported and used for simulating particle collision for two crystals entering an optical trap.

\subsubsection*{Electrostatic Forces}

The electrostatic force corresponds to a force (typically repulsive) between colloidal particles due to surface potentials and corresponding responses of ionic environment. Approximating our crystals as spheres, the magnitude of electrostatic force between two identical crystals can be described as\cite{Hogg1966,Ohshima1994}:
\begin{equation}
    |\textbf{F}_{elec}| = 2\pi\epsilon_r\epsilon_0a\psi_s^2\kappa_De^{-\kappa_Dh}(1-e^{-\kappa_Dh})
    \label{Felec}
\end{equation}
where $\epsilon_r$ and $\epsilon_0$ are, respectively, the permittivity of the relative electrolyte solution and the permittivity of vacuum, $\psi_s$ is the surface potential of the spheres, and $\kappa_D$ is the inverse of the Debye Length.
Equation \ref{Felec} is valid when $\kappa_Da \gg 1$ and $h \ll a$ that are satisfactory to cover the length scale of interest for self-assembly based on the size of particles in our study ($a=500$ nm and 62.5 nm).
To approximate the surface potential, the zeta potential of the crystals suspended in solution were measured. The zeta potential for the rough 1 $\mu$m diameter $\alpha$-NAYF crystals was 29.1 $\pm$ 4.0 mV, although it varied slightly from synthesis to synthesis. Similarly, the zeta potential of the smooth $\alpha$-NAYF crystals ($\sim$ 125 nm in diameter) was 34.3 $\pm$ 2.4 mV, indicating no significant difference compared to that of the rough 1 $\mu$m diameter $\alpha$-NAYF crystals.

\subsubsection*{van der Waals Forces}

Lastly, the van der Waals force is a major attractive force for colloidal particles at relatively short separation distances. Similar to the electrostatic force, we approximated our crystals as spheres, which leads to the magnitude of the van der Waals force between two identical crystals as\cite{Pailthorpe1982}:
\begin{widetext}
\begin{equation}
    |\textbf{F}_{vdw}| = -\frac{A_H}{3}\left[ \frac{2a^2(h+2a)}{[(h+2a)^2-4a^2]^2}+\frac{2a^2(h+2a)}{[(h+2a)^2]^2}-\frac{h+2a}{(h+2a)^2-4a^2}+\frac{h+2a}{(h+2a)^2} \right]
    \label{FirstFvdw}
\end{equation}
\end{widetext}
where $A_H$ is the Hamaker constant for the material of interest. At $h\ll a$, Equation \ref{FirstFvdw} can be further simplified to:
\begin{equation}
    |\textbf{F}_{vdw}| = -A_Ha/h^2
\end{equation}
While the Hamaker constant can be rigorously obtained from frequency-dependent dielectric functions of the system's constituents, we used the following approximate equation due to unavailable data for the dielectric functions\cite{Israel6}:
 \begin{equation}
    A_H = \frac{3}{4}kT\left(\frac{\epsilon_1-\epsilon_3}{\epsilon_1+\epsilon_3}\right)^2+\frac{3\hat{h}\nu_e(n_1^2-n_3^2)^2}{16\sqrt{2}(n_1^2+n_3^2)^{3/2}}
    \label{HamakerEq}
\end{equation}
 where $\epsilon_1$ and $n_1$ denote dielectric constant and refractive index of $\alpha$-NAYF (8.16 and 1.47 respectively)\cite{mpNAYF}, and $\epsilon_3$ and $n_3$ denote dielectric constant and refractive index of water (78.5 and 1.33 respectively)\cite{Ninham1970}. The characteristic electronic absorption frequency $\nu_e$ is assumed to be $2.0\times 10^{16}$ [rad/s]\cite{Ninham1970} and $\hat{h} =6.6256\times 10^{-34}/2\pi$ [J$\cdot$s/rad]. Note that Equation \ref{HamakerEq} is based on an assumption that the frequency-dependent dielectric functions can be represented by UV absorption frequency and refractive index for nonzero frequency contributions. Using all relevant numerical data, the Hamaker constant of our system is 2.4465 $kT$ or $1.0066\times 10^{-20}$ J at 298.15K.

\subsection*{Modeling the Surface Roughness of Spherical Particle}

To calculate hydrodynamic forces between two rough spheres, we first needed a way to model the rough-surfaced spherical particles. We numerically constructed a rough sphere by using spherical meshing (based on Gmsh\cite{gmsh}) where elements of the sphere are allowed to randomly reduce the distance from the center of the sphere. We used a boundary element method that implements an integral representation of the Stokes flow via a hydrodynamic Green’s function (i.e., the Oseen tensor), coupled with a scheme to generate surface meshes for describing particles, called BEST\cite{Aragon2004,Aragon2006}. We further developed a simple model to construct the surface roughness by allowing mesh elements of the spherical particle to randomly reduce the distance from the center of the sphere over random azimuthal and polar positions of the original smooth sphere. 

\begin{figure*}
\includegraphics[width = 16cm]{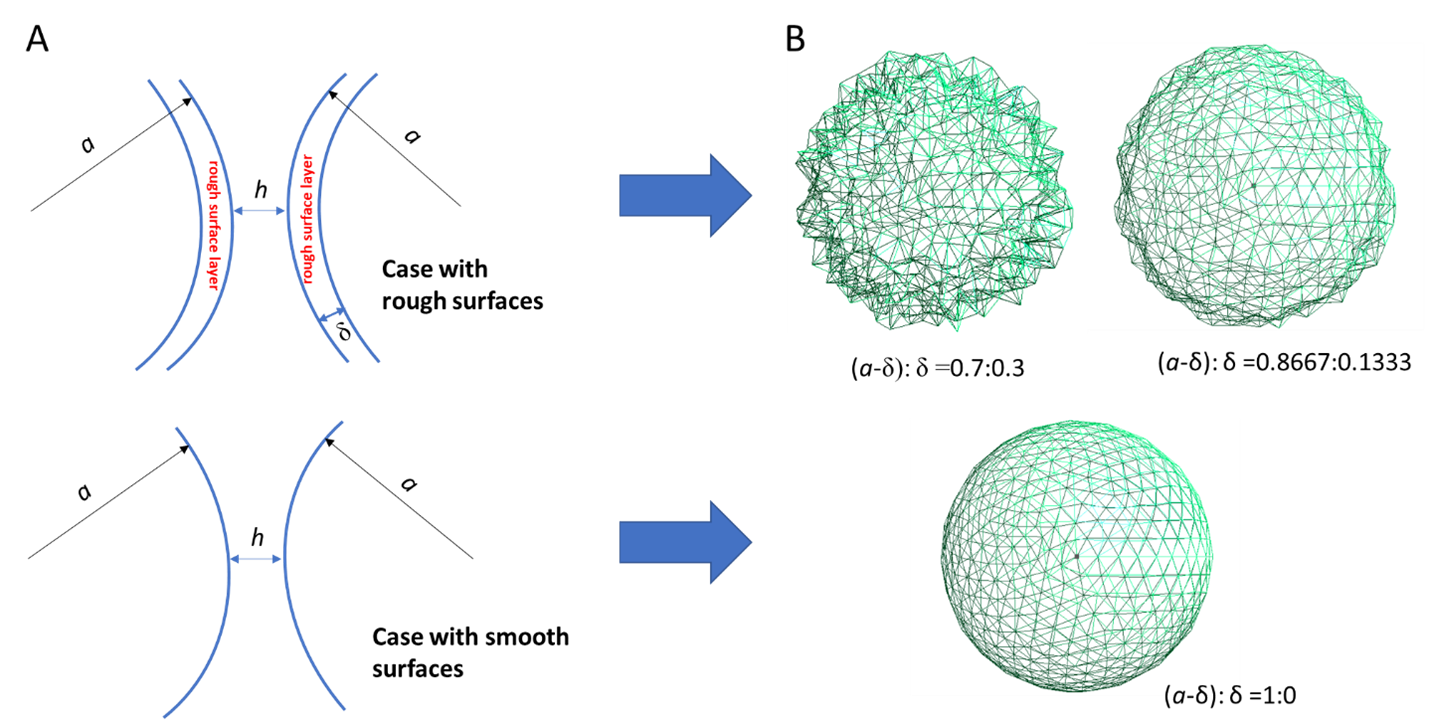}
\caption{(A) Schematic diagrams for modeling the surface roughness (B) Constructs of two different levels of surface roughness in comparison to the smooth sphere. Here, the separation distance $h$ in (A) denotes a separation between the particles. For these example objects, 634 mesh elements were used.} 
\label{Spheres}
\end{figure*}

As shown in Fig. \ref{Spheres}, we first defined a “rough surface layer” by defining a maximum radius deviation ($\delta$), in comparison to the particle radius ($a$), to be consistent with relevant definitions (e.g., radius and the separation between particles) for the smooth case. This is also consistent with the construction shown in a previous study by Jenkins \& Koenders\cite{Jenkins2005}. At a given $\delta$, we used a random number between 0 to 1 to realize the surface roughness within the rough surface layer. The radius $a$ is then modified by the random number generated using $\delta$. The total roughness can be described by the ratio of the minimum possible radius to the maximum possible radial deviation, $(a-\delta)$:$\delta$. Fig. \ref{Spheres} illustrated two rough cases for our study: 1) $(a-\delta)$:$\delta$ = 0.7:0.3 (strong roughness) and 2) $(a-\delta)$:$\delta$ = 0.8667:0.1333 (weak roughness). Note that the strong roughness better represents the $\alpha$-NAYF particles used in the experiment. The smooth case corresponds to $(a-\delta)$:$\delta$ = 1.0:0.0.

\subsection*{Details on hydrodynamic resistivity calculations}

\begin{figure*}
\includegraphics[width = 16cm]{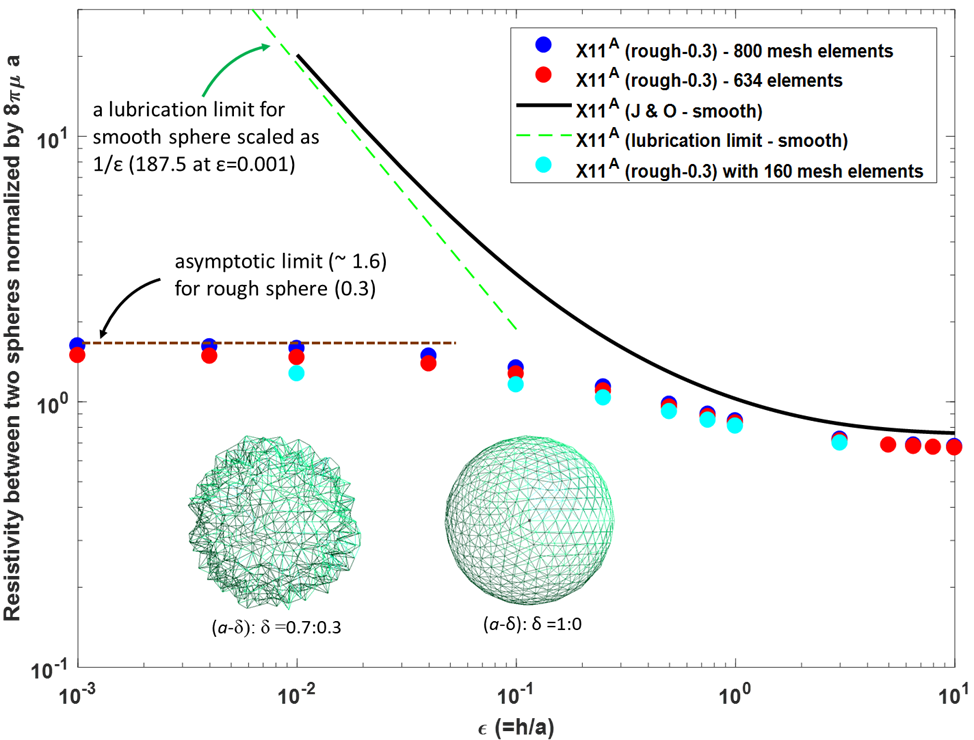}
\caption{Hydrodynamic resistivity between two spheres for both smooth and rough cases, scaled by $8\pi\mu a$ where $\mu$ and $a$ denote the viscosity of surrounding fluid and the particle radius respectively. Here, $a$ is the particle radius and a normalized separation, $\epsilon$, is defined by a separation between spheres ($h$) divided by $a$. The strong roughness case, denoted as “rough-0.3” (($a-\delta$):$\delta$ = 0.7:0.3), were used and three different mesh elements (800, 634, 160) for describing the particle were used for comparison. The “J \& O” and “lubrication limit” denote numerical results and asymptotic form at lubrication regime from Jeffrey and Onishi\cite{Jeffrey1984}.} 
\label{HydDyn}
\end{figure*}

We calculated hydrodynamic forces between two rough spheres by using a boundary element method that employs an integral representation of the Stokes flow via the Oseen tensor called BEST\cite{Aragon2004,Aragon2006}, combined with a mesh representation of the particle as described above. Fig. \ref{HydDyn}, represented by a normalized hydrodynamic resistivity by $8\pi\mu a$ where $\mu$ is the viscosity of surrounding fluid, showed our calculations depending on three different mesh elements (800, 634, 160) for the particles, along with other known results for smooth spherical particles, “J-O” and “lubrication limit”. It is important to note that hydrodynamic resistivity was normalized by $6\pi\mu a$ in Fig. \ref{forces} in the main text to clearly represent the Stokes’ drag at large separations, causing it to approach unity after normalization. Because of inherent nature involved in the integral representation, we examined the effect of mesh elements on the hydrodynamic resistivity. The calculations clearly indicate that the number of mesh elements does not influence the hydrodynamic resistivity. An insignificant increase in the hydrodynamic resistivity from 160 to 634 and finally to 800 mesh elements suggests that 800 mesh elements would be sufficient for calculations, which was shown in Fig. \ref{forces} in the the main text. For all mesh element cases, the hydrodynamic resistivity approaches to $\sim$ 1.6 as indicated in Fig. \ref{HydDyn}, corresponding to $\sim$ 2.1 when the resistivity was scaled by $6\pi\mu a$ due to the different normalization. As clearly shown, the roughness drastically reduces the hydrodynamic resistivity or hydrodynamic forces between two particles.

\begin{figure*}
\includegraphics[width = 16cm]{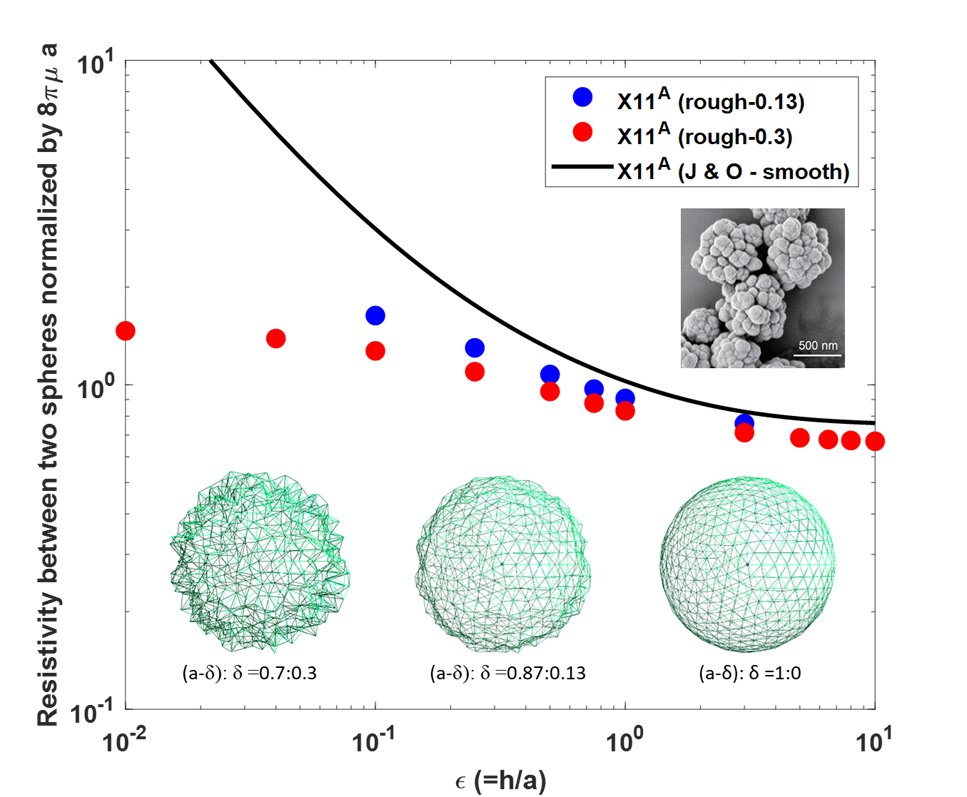}
\caption{Hydrodynamic resistivity between two spheres in smooth and rough cases, scaled by $8\pi\mu a$ where $\mu$ and $a$ denote the viscosity of surrounding fluid and the particle radius respectively. Here, $a$ is the particle radius and a normalized separation, $\epsilon$, is defined by a separation between spheres ($h$) divided by $a$. The strong roughness case, denoted as “rough-0.3” ($(a-\delta)$:$\delta$ = 0.7:0.3), and the weak roughness case, denoted as “rough-0.13” ($(a-\delta)$:$\delta$ = 0.8667:0.1333), were used. The “J \& O” data are also shown to compare with the smooth case. The insets are 3D contructions of spheres corresponding to two different levels of surface roughness in comparison to the smooth case, along with the image of $\alpha$-NAYF particles.} 
\label{Meshes}
\end{figure*}

Although the strong roughness case ($(a-\delta)$:$\delta$ = 0.7:0.3) is a much better representation of our $\alpha$-NAYF particles shown in Fig. \ref{NAYF}, we also examined a different level of the surface roughness, $(a-\delta)$:$\delta$ = 0.8667:0.1333 (weak roughness), based on 634 elements. Fig. \ref{Meshes} indicates that the different roughness level would change the reduction of the hydrodynamic resistivity. However, the change is relatively small in comparison to the smooth case at the lubrication regime, $\epsilon \ll$ 1. 

\subsection*{Langevin Dynamic Simulations and Data Analysis}

\begin{figure*}
\centering
\includegraphics[width = 12cm]{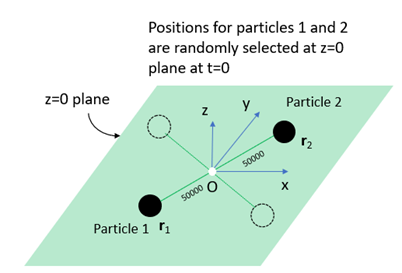}
\caption{A schematic for the Langevin dynamics simulations in our study. } 
\label{LangevinScheme}
\end{figure*}

Simulations for a pair of particles have been performed to understand the self-assembly of $\alpha$-NAYF particles based on the Langevin equation with negligible particle inertia, called the mobility equation\cite{Russel1989} (Equation \ref{velAllForces} in the main text), coupled with all relevant forces and hydrodynamic mobility. The simulation starts with initial positions of two particles that ensure a sufficient separation between them, especially considering significant inward attractive forces caused by the optical tweezers. Specifically, at the initial step, the centers of two particles are placed at z = 0 plane and the distance ($|\textbf{r}_1-\textbf{r}_2|$) is set to 100000 that is equivalent to 10.5$a$ or $h$ = 8.5$a$, as shown in Fig. \ref{LangevinScheme}. At each time step, ($2.42\times10^{-2} \mu$s for $a$ = 500 nm; $2.42\times10^{-4} \mu$s for $a$ = 62.5 nm), the positions of two particles are updated by solving Equation \ref{vel:force} numerically to check if both particles would come into contact. Note that the time steps are less than the viscous relaxation time of particles to ensure that the fluctuation-dissipation theorem for Brownian forces ($\textbf{F}_B$), $\langle\textbf{F}_B(t)\textbf{F}_B(0)\rangle =12\pi\mu akT\textbf{K}\delta(t)$, can be rigorously applied\cite{Russel1989}. Each simulation typically spans up to $10^7-10^8$ time steps that correspond to more than $10^7-10^8$ times of the viscous relaxation times of particles) to avoid any missing possible collision/contact events. All collision events and corresponding times required for the contact/collision are counted over 8000 independent ensembles to ensure enough statistics, based on independent initial positions and Gaussian random numbers for realizing Brownian forces at every time step. Finally, the collision times are binned, and the ensemble-averaged results are represented as probability density function, as shown in Fig. \ref{pdfs} in the main text.

\clearpage
\beginsupplement
\section*{Supplementary Figures}

\begin{figure*}
\includegraphics[width = 10cm]{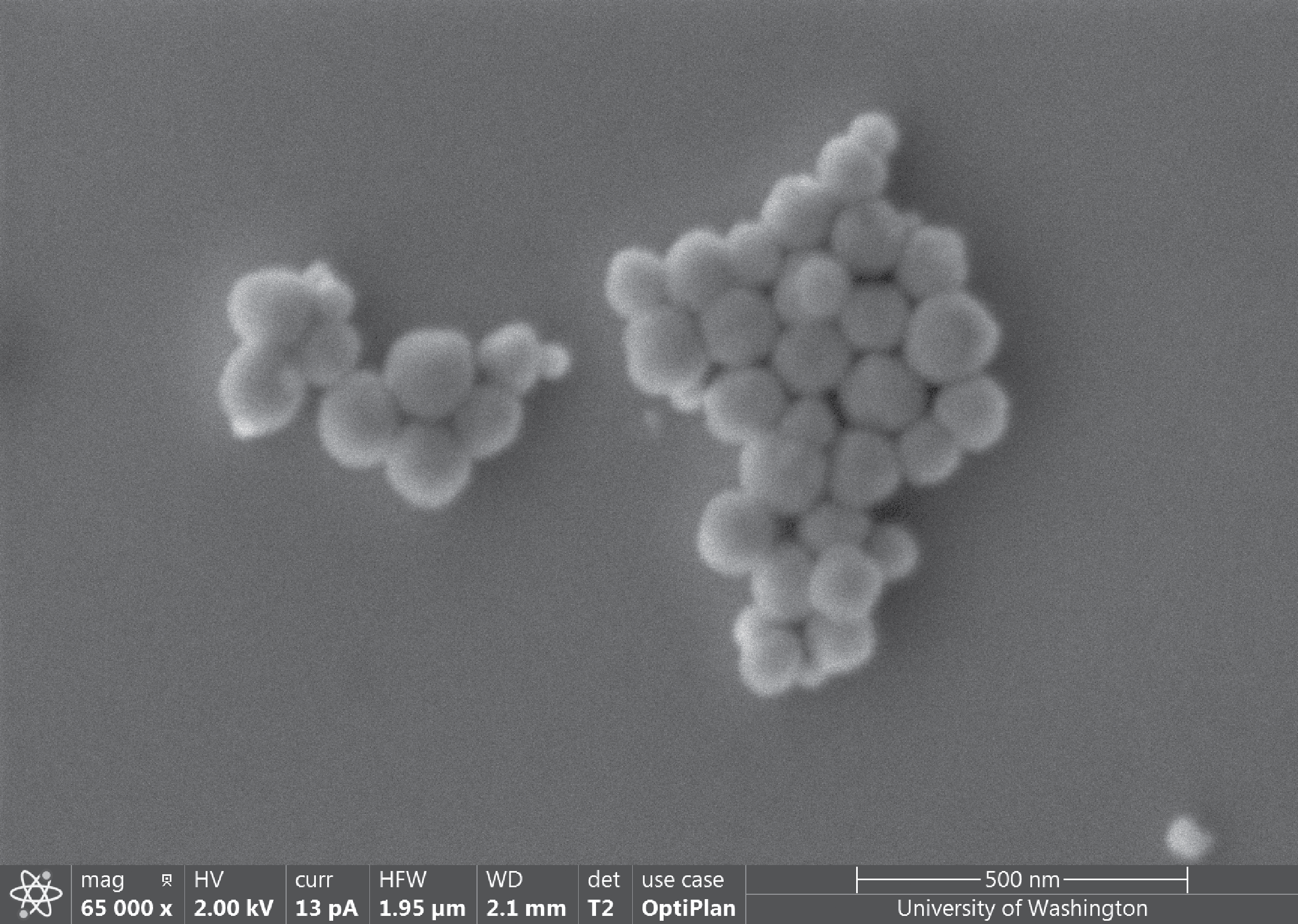}
\centering
\caption{SEM of the small, smooth $\alpha$-NAYF used in the experiments. The surfaces are smoothed by synthesizing in the presence of EDTA for use as a ligand. The surface is smooth in contrast to the rough particles, and we approximate them as spheres in the calculations presented in this work.}
\label{smooth}
\end{figure*}

\clearpage

\section*{Supplementary Movies}
\renewcommand{\figurename}{Supplementary Movie}
\setcounter{figure}{0} 

\begin{figure*}
\centering
\caption{[AssemblyVideo.mp4] \quad This video file can be downloaded separately as part of the supplementary materials. A video of the assembly process as captured using a scientific camera collecting light through the trapping objective. A sample is trapped in the laser, which is located in the white circle. After trapping a first particle, several more are brought into the trap and allowed to attach spontaneously. Eventually, a final crystal is added, which attachs to the side of the existing chain. 50 seconds into the video, the laser is turned off and the chain of $\alpha$-NAYF begins to fall to the bottom of the trapping chamber. The camera focus is manually adjusted to see the chain as it falls.}
\label{video}
\end{figure*}

\begin{figure*}
\caption{[Animation.mp4] \quad This video file can be downloaded separately as part of the supplementary materials. An animation of the assembly process to demonstrate the generally process with more clarity. Rough particles of $\alpha$-NAYF are floating in Brownian, then drawn individually into the trap. They assemble into a chain in the trap with the last attaching to the side, as seen in Supplementary Movie \ref{video}. The animation is not drawn to scale and the motion of the crystals is demonstrative only and is also not to scale.}
\label{animation}
\end{figure*}

\end{document}